\documentstyle{aipproc}

\newcommand{\be}{\begin{equation}}
\newcommand{\ee}{\end{equation}}
\def\vh{\varphi}

\begin{document}
\title{Conformal Symmetry and Unification\thanks{Talk given at the International 
Conference {\it Particles, Field and Gravitation}, Lodz, April 1998}}
\author{Marek Paw\l owski\thanks{e-mail: 
pawlowsk@fuw.edu.pl.}}
\address{Soltan Institute for Nuclear Studies, Warsaw, Poland}

\maketitle

\begin{abstract}

The Weyl-Weinberg-Salam model is presented. It is based on the local conformal 
gauge symmetry. The model identifies the Higgs scalar field in SM with the 
Penrose-Chernikov-Tagirov scalar field of the conformal theory of gravity. Higgs 
mechanism for generation of particle masses is replaced by the originated in 
Weyl''s ideas conformal gauge scale fixing. Scalar field is no longer a dynamical 
field of the model and does not lead to quantum particle-like excitations that 
could be observed in HE experiments. Cosmological constant is naturally 
generated by the scalar quadric term. The model admits Weyl vector bosons that
can mix with photon and weak bosons.

\end{abstract}

\section*{Introduction}

In 1918, Herman Weyl presented the idea and notion of gauge invariance 
\cite{weyl}. It was a consequence of natural generalization of Riemannian 
geometry used in Einstein's General Relativity theory (GR). Weyl assumed that 
Einstein's metricity condition
\be
\nabla g=0
\label{metricity}
\ee
could be replaced by a less restrictive conformal condition
\be
\nabla g_{\mu\nu}\sim g_{\mu\nu}.
\label{weyl}
\ee
Thus he supposed that for a vector transported around a closed loop by parallel 
displacement not only the direction but also the length can change, but the 
angle between two parallelly transported vectors has to be conserved. Weyl 
observed that if the Einstein's torsion free condition
\be
\Gamma^{\lambda}_{\mu\nu}-\Gamma^{\lambda}_{\nu\mu}=T^{\lambda}_{\mu\nu}=0.
\label{torsion}
\ee
is kept, then - similarly as in the case of GR - there is a relation between the 
metric and the affine structure of tangent boundle $TM$. In contrary to GR case, 
the Weyl connection is not given uniquely by the Christophel symbol: it could 
depend on an arbitrary vector field in principle. This vector field is a 
compensating potential for a local conformal group of scalar multiplicative 
transformations conserving the conformal condition (\ref{weyl}). Weyl called 
this group the gauge group as it sets a reference scale from point to point in 
the space-time. Initially he interpreted the new vector field as the 
electromagnetic potential and has proposed a dynamics for the model that was 
based on the bilinear in the generalized curvature Lagrangian. 

The dynamics of original Weyl's theory turned out to be much more complicated 
than the dynamics of Einstein GR. The idea of gauge conformal invariance of the 
theory was also a subject of intensive criticism. Weyl's conformal theory leaves 
the freedom for the space-time dependent choice of length standards. It seamed 
that this gauge freedom clashes with quantum phenomena that provide an absolute 
standard of length. The point is however, that the freedom to set arbitrary 
length standards along an atomic path does not mean that atomic frequencies will 
depend on atomic histories (what was the most popular argument in early 
literature). In Weyl's theory, an atomic frequency depends on the length 
standard at a given point but not on a history of the atom. Simultaneously, all 
other dimensional quantities measured at this point depend on this standard in 
the same way. Consequently dimensionless ratios are standard independent and 
experimental predictions do not depend on a particular conformal gauge fixing.

The more fundamental arguments raised against conformal theory were based on the 
reasonable claim that an acceptable theory should not introduce needles objects 
and notions. If atomic clocks measure time in an absolute way and velocity of 
light is an absolute physical quantity (or is definite at least) then the 
relativism of length is unnatural and redundant. However, we should point out a 
very essential assumption concerning atomic clocks that is hidden in the above. 
This is an extra\-po\-lation of our {\sl flat and first order experience} that 
all atomic clocks are proportional always and everywhere. One assumes -- roughly 
speaking -- that {\sl the ratios of electron mass to proton mass and to other 
quantum standards are always and everywhere the same}. One can believe that this 
statement is true but one should remember (especially when such effects like red 
shift or other distant signals are interpreted) that at the large scale this 
statement is only an assumption. It should be (and it could be! 
\cite{constants}) a subject of experimental verification. Conformally invariant 
gauge theory apparently makes a room to relax from such {\sl a priori} 
suppositions \cite{dirac73}, but in fact it does not predicts itself a dynamics  
for evolution of fundamental physical ""constants""

\smallskip

Weyl conformal theory is a gauge theory of length. It was proposed as a 
geometrical theory of electromagnetism. Soon after, Dirac proposed his theory of 
quantum relativistic electron in the flat space \cite{dirac28}. It was a gauge 
theory of complex electron's phase and it turned out that it provides more 
adequate framework for description of electromagnetic phenomena. Weyl's proposal 
was abandoned by the author himself (but still in 1973 Weyl's gauge theory of 
scale was considered by ... Dirac as a candidate for description of 
electromagnetism \cite{dirac73}). 

The original Dirac's theory of electron was extended to the curved space case 
\cite{schouten,schroedinger,infeld}. Taking a four-dimensional manifold $M$, a 
copy of two dimensional complex vector field $F_pM$ can be attached to every 
point $p$ of $M$. Then two, in principle independent pairs of affine and metric 
structures can be implemented on the manifold. The natural tangent boundle $TM$ 
can be equipped with an affine connection $\Gamma$ and the field of metric $g$. 
Independently a connection $\gamma$ can be defined in the boundle $FM$ and an 
arbitrary field $\varepsilon$ of Levi-Civita metric can be chosen (for generic 
two-dimensional complex vector space there is a natural class of antisymmetric 
Levi-Civita metrics that differ by a complex factor). 

The two structures $\{\Gamma, g\}$ and $\{\gamma, \varepsilon\}$ can be 
naturally correlated. The important observation is that the Levi-Civita metric  
$\varepsilon$
induce Lotentz metric $\varepsilon\otimes\overline{\varepsilon}$  
at
every fiber of the tensor product boundle $FM\otimes \overline{FM}$ (see e.g. 
\cite{wald} for further details). Thus the real part of $FM\otimes  
\overline{FM}$
(which is a four dimensional real vector boundle) can be related with the 
tangent
vector boundle $TM$.
 
It was found by Infeld and van der Waerden \cite{infeld} that such correlation 
of boundles correlates also their metrics and affine structures. Keeping the 
restrictions of GR (metricity and torsion-free) they have shown that metric 
structure $\varepsilon$ of $FM$ is given by metric structure $g$ of $TM$ up to 
the arbitrary phase factor. Simultaneously the affine structure $\gamma$ of $FM$ 
is given by the affine structure
$\Gamma$ of $TM$ up to an arbitrary vector field. This new vector field is a 
compensating potential for the $U(1)$ local symmetry group of phase 
transformations of all Dirac fields in the theory. The authors have identified 
this new field with electromagnetic potential. Such identification was a subject 
of criticism as the new vector potential has been coupled universally to all 
fermions including chargeless neutrino. The modern Weinberg-Salam theory (WS) 
predicts that all fermions couple to $U(1)$ gauge field. There is a second 
nonabelian gauge group $SU(2)$ in the theory acting only on the left components 
of Dirac bispinors. Due to the structure of couplings and the effective mass 
matrix for gauge bosons the massless field - naturally identified with photon - 
is a combination of original $U(1)$ and $SU(2)$ bosons. It does not couple to 
neutrinos despite the fact that the original abelian vector potential does. Thus 
the Infeld - van der Waerden vector potential can be naturally identified with 
$U(1)$ gauge group potential of the WS model without any conflict with theory 
and experiment. 

The rest of the present paper is devoted to the description of the version of 
Weinberg-Salam theory conformally coupled with Weyl's theory of gravity. The 
first version of the model was proposed in \cite{foundations} (see also 
\cite{gpps}). Similar ideas were also presented in \cite{cheng}. More 
comprehensive list of the bibliography of the subject can be found in 
\cite{hehl}.

Taking into account the roots of the theory it could be called the Weyl-
Weinberg-Salam model (WWS).

\section*{Weyl-Weinberg-Salam model}

Let us fix the notation

Weyl's potential will be denoted by $S_{\mu}$.  Let us assume torsion free 
condition (\ref{torsion}).
Then the connection in $TM$ is given by

\be
\Gamma^{\rho}_{\mu\nu}=\{^{\rho}_{\mu\nu}\}+
f(S_{mu}g^{\rho}_{\nu}+S_{\nu}g^{\rho}_{\mu}-S^{\rho}g_{\mu\nu})
\label{connection}
\ee
where $f$ is an arbitrary coupling constant  (in principle it could be absorbed 
at this level by a redefinition of $S_{\mu}$ but it is convenient to keep it 
here and set its value later). Consequently Weyl's conformal condition 
(\ref{weyl}) gets the form
\be
\nabla_{\mu} \hat{g} = -2fS_{\mu}\hat{g}
\label{weyl2}
\ee
Equations (\ref{connection}) and (\ref{weyl2}) are invariant  
with
respect to Weyl''s transformations
\be
g_{\mu\nu}\to \Omega^2 g_{\mu\nu}=e^{2\lambda} g_{\mu\nu}
\label{transmetric}
\ee
\be
S_{\mu}\to S_{\mu}-{1\over f}\partial_{\mu}\lambda.
\label{transweyl}
\ee
Thus metric tensor is covariant with respect to Weyl''s transformations with 
degree 2. The Riemann and Ricci tensors constructed from (\ref{connection}) are 
conformally invariant objects but their contraction to scalar curvature $R$ is 
not. $R$ can enter linearly to a conformally invariant expression of dimension 
of action if it is combined with a scalar Penrose-Chernikov-Tagirov (PCT) field  
$\vh_{_{PCT}}$ \cite{PCT} that transforms according to 
\be
\vh_{_{PCT}} \to e^{-\lambda}\vh_{_{PCT}}.
\label{transphi}
\ee
Then the combination $\vh_{_{PCT}}^2R$ is conformally invariant. The conformal 
covariant derivative of $\vh_{_{PCT}}$ is given by
\be
\nabla_{\mu}\vh_{_{PCT}} = (\partial_\mu -fS_{\mu})\vh_{_{PCT}}
\label{nablaphi}
\ee
and it transforms according to (\ref{transphi}).

The most general conformally invariant Lagrangian that leads to second order 
equations of motion for the metric-Weyl-scalar system reads \cite{padman}
\be
L_g= -{\alpha_1\over 12}\vh_{_{PCT}}^2 R + {\alpha_2\over 2}\nabla_{\mu}
\vh_{_{PCT}}\nabla^{\mu}\vh_{_{PCT}} - {\alpha_3\over 4}
H_{\mu\nu}H^{\mu\nu} -{\lambda\over 4{\rm !}}\vh_{_{PCT}}^4
\label{Lg}
\ee
where
\be
H_{\mu\nu}=\partial_{\mu}S_{\nu}-\partial_{\nu}S_{\mu}.
\label{Hmunu}
\ee
The coupling constants $\alpha_1$, $\alpha_2$ and $\alpha_3$ are arbitrary but 
the last two constants can be absorbed in $\vh_{_{PCT}}$ and $S_{\mu}$ by a 
suitable redefinition of the fields. Observe however, that we are not able to 
absorb simultaneously $\alpha_3$ and $f$. The last coupling remains arbitrary 
and has to be fixed by experiment. 

We can also include the original Weyl Lagrangian being the square of Weyl tensor 
$L_W=\rho C^2$ where $\rho$ is a coupling constant.
 
Now we can face the Weinberg-Salam part, or more generally, the full Standard 
Model of fundamental interactions \cite{rpp}. 

First, we should recall \cite{hayashi} that Weyl's vector potential $S_{\mu}$ do 
not couple directly to Dirac fermions if they transforms according to the rule
\be
\Psi\to e^{-{3\over 2}\lambda}\Psi.
\label{transpsi}
\ee

The conformally invariant part of SM can be written in the following form:
\be
\label{smc}
L_{SM}^c[\vh_{H},{\bf n},V,\psi,g] ={\cal
L}_0^{SM}+ [-\vh_{H}F+\vh_{H}^2B-\lambda\vh_{H}^4].
\ee
${\cal L}_0^{SM}$ is the conventional SM Lagrangian without the ``free" part for 
the modulus of the Higgs $SU(2)$ doublet $\vh_H$ and without the Higgs mass 
term; $B$ is the mass term of the vector fields generally denoted by $V$ and $F$ 
is the mass terms of the spinor fields generally denoted by $\psi $
\be
\label{66}
B=D{\bf n}(D{\bf n})^*\,;\,F=(\bar\psi_L{\bf n})\psi_R+ h. c.;~~~~~
{\bf n}=\left(\begin{array}{c} {\rm n}_1 \\
{\rm n}_2 \end{array}\right);\;\;
{\rm n}_1\stackrel{*}{{\rm n}_1}+n_2\stackrel{*}{{\rm n}_2}=1;
\ee
${\bf n}$ is the angular component of the Higgs $SU(2)$ doublet.

As there are two abelian gauge groups in the model also a mixed term 
\be
L_{SB}=\alpha_4 H_{\mu\nu}F^{\mu\nu }
\label{LSB}
\ee
is admitted by all symmetries of the model in general.

The main idea of conformal unification consists in the identification of PCT 
scalar field $\vh_{_{PCT}}$ with the modulus of Higgs doublet $\vh_H$ within the 
rescaling factor $\chi$
\be \label{67a}
\vh_H=\chi\vh_{_{PCT}}.
\ee

The total lagrangian of the conformally unified WWS model can be written as a 
sum of three terms described above
\be
L_T=L_g+L_{SM}^c+L_{SB}
\label{LT}
\ee
with the constraint (\ref{67a}) resolved. 

The rescaling factor $\chi$ of (\ref{67a}) is a new coupling constant, which 
coordinates weak and gravitational scales \cite{part1}.

\section*{Scale fixing}

The theory given by (\ref{LT}) does not contain any dimensional parameter. This 
is the necessary condition for it to be conformally invariant. As it was 
discussed in the Introduction in the context of the Weyl theory alone, 
dimensional quantities are observed in nature only indirectly. Measuring one of 
them, we always refer to some other dimensional quantity. We measure ratios of 
dimensional quantities and we are not able to measure anything more. Our 
statements express the ratios in the form that carries in the content of its 
measure an information on the denominator. Thus the dimensional quantities in 
the half seams to be nothing but only a product of human invention, a logical 
and a lingual abbreviation representing both the physical information and the 
chosen convention. There is no doubt that the abbreviation is convenient and 
useful in practice - in our "flat" surrounding at least (see however the 
Introduction again). The conformal theory reproduces this conventional 
abbreviation. It could be done with the help of the most natural mechanism for 
this purpose, the mechanism of scale fixing which is an example of the gauge 
fixing of the conformal gauge symmetry group (it is in fact the first historical 
example of the notion of gauge).

Gauge fixing freedom allows us to impose an additional condition on the theory 
variables. All lawful conditions (those that can be fulfilled by the fields 
obtained from a generic configuration by a gauge symmetry transformation) are 
classically equivalent but not all of them are equally convenient for a given 
practical purpose. In the case of our conformal theory, we are free to fix the 
dimensional scale. A natural choice is the one that fixes particle masses in our 
flat surrounding to theirs conventional space-time independent values. (In fact, 
nobody will admit in practice that the choice could be a different.) This could 
be achieved for the conformal symmetry gauge condition that fixes the scalar 
field in $L_{SM}^c$ (\ref{smc}) to a constant (space-time independent) value. 
Thus we can demand that
\be
\label{fix}
\vh_H=const=v
\ee
and it is clear that a generic nonzero scalar field configuration can be 
conformally transformed to fulfill condition (\ref{fix}).

Choosing $v=246 \rm GeV$ and choosing ordinary unitary gauge of weak group, we 
reproduce the whole structure of classical SM masses in WWS model.

\smallskip

It should be stressed here that no mechanism of spontaneous or dynamical 
symmetry breaking was used in order to produce particle masses. The conformal 
gauge fixing condition (\ref{fix}) was a sufficient tool. Let us also comment - 
but without further discussion - that however the condition (\ref{fix}) serves 
for easy identification the flat space particle content of the model, it needn't 
be the best toll for other purposes. The condition leads to a massive sigma 
model that is not perturbatively renormalizable. (The fact does not prejudge the 
renormalizability of the theory - if we can speak at all about a 
renormalizability of the theory including gravity. A convenient choice of gauge 
fixing condition is essential for perturbative analysis of the renormalizability 
problem. It is known, e.g. that the unitary gauge is not the best choice for 
this task in SM.) 

\section*{Toward experiment}

The properties of theory given by (\ref{LT}) depend on the value of coupling 
constants $\alpha_i$, $\rho$, $f$, $\lambda$ and $\chi$. 

The striking feature of the conformal theory is the lack of ordinary Einstein 
term in (\ref{LT}). Observe however, that even in the simplest case $\chi=1$ 
(the Higgs field identified with PCT scalar field), the condition (\ref{fix}) 
allows us to reproduce easily the Einstein term \cite{cheng}. It is sufficient 
to demand that 
\be
-{\alpha_1\over 12}v^2={1\over 8\pi G}
\label{newton}
\ee

If the conformal gauge fixing condition (\ref{fix}) is chosen, a mass term for 
the Weyl's vector field $S_{\mu}$ appears \cite{cheng} and $S_{\mu}$
acquires mas
\be
m_S^2={1\over 2}f(\alpha_2-\alpha_1)v^2
\label{ms}
\ee
The condition (\ref{newton}) leads to Weyl vector mass
\be
m_S=0.5\cdot 10^{19} f\cdot GeV.
\ee
In turn the Weyl''s mass equals zero only in the special case when 
$\alpha_2=\alpha_1$. Then an additional symmetry is realized in the model.
Without changing the action we can transform according to the rules of conformal 
transformations (\ref{transmetric}), (\ref{transphi}) and (\ref{transpsi}) the 
metric, the scalar and the all fermion fields leaving the Weyl field unchanged. 
Similarly we can independently transform $S_{\mu}$ and (\ref{LT}) will not 
change. In that case the Weyl potential decouples from scalar field and if 
$\alpha_4=0$, it is coupled only to gravity. 
We get Penrose-Chernikov-Tagirov theory of scalar field conformally coupled with 
gravity \cite{PCT}. In order to reproduce appropriate Newtonian limit already at 
the classical level, we have to demand that $\chi$ is very 
small\cite{foundations,part1}
\be
\chi\sim {v \over m_{_{PLANCK}}}
\ee

In the flat limit approximation (the condition (\ref{fix}) is applied, dynamics 
of $g$ is frozen and $g$ is chosen to be the metric of Minkowski space) the 
conformally unified WWS theory leads to the SM-like $\sigma$-model. ( It holds 
independently on the values of couplings $\alpha_i$, $\rho$, $f$ and $\lambda$ 
in (\ref{LT})). There is still $U(1)\times SU(2)_L\times SU(3)$ gauge symmetry 
but the feature of perturbative renormalizability is lost. Despite this fact the 
theory is still predictive. We can reproduce all SM 1-loop results for the 
processes without external Higgs lines. The SM Higgs mass is replaced in 
calculations by an effective cutoff that can be expressed (eliminated) by some 
measured quantity or a combination of observables. 1-loop predictions for 12 LEP 
observables were given in \cite{part2} in reasonable agreement with SM and 
experiment.

The flat limit of the presented unified WWS model can be a subject of 
experimental verification and discrimination. The direct verification will be 
provided (of course!) by LHC. This installation should produce data able to 
cover all admissible SM Higgs mass range. If no Higgs signal will be found (and 
we know from LEP that it should be found there if SM is valid) then conformal 
unified model predicting no dynamical scalar particle at all should be a serious 
alternative. In turn founding at LHC Higgs particle with the all its SM 
predicted properties will tell us that the minimal conformal unification is not 
good. 

There was also proposed an indirect method for verification of the flat limit 
consequences of WWS model \cite{mu}. It is based on the observation that while 
the SM Higgs mass $m_H$ is an energy independent physical constant, the cutoff 
$\Lambda$ introduced in the 1-loop analysis of $\sigma$-model can depend in 
principle on the energy of the process considered and on its other parameters. 
The idea is to derive $m_H$ from two experiments performed at different energy 
scales. If it will happen that the derived masses disagree it will mean that SM 
fails while WWS accepts this phenomenon. This is a kind of negative test of SM. 
It was estimated that the proposed comparison could be made on the base of data 
given by LEP and CESR B and PEP II if some demanded but realistic luminosity 
will be achieved. 

\section*{Conclusions}

The Weyl-Weinberg-Salam model identifies the Higgs scalar field in SM with the 
Penrose-Chernikov-Tagirov scalar field of the conformally invariant theory of 
gravity. This identification is very natural and it leads to important physical 
consequences:

Higgs mechanism for generation of particle masses is replaced by the originated 
in Weyl's ideas conformal gauge scale fixing. Scalar field is no longer a 
dynamical field of the model -- it is rather a Goldsone direction in field 
space, the direction that is tangent to the conformal gauge group. Consequently 
it does not lead to quantum particle-like excitations that could be observed in 
HE experiments and it does not acquire quantum expectation value in the vacuum. 
Experimental flat limit consequences of the model could be tested in near 
future.

No cosmological consequences characteristic for the SM Higgs field can be 
derived from the present model, but the scalar sector generates cosmological 
consequences in a different way. The quadric coupling constant $\lambda$ of 
thescalar PCT field which in WWS does not play any role in generating particle 
masses, has its effect in generation of cosmological constant. This constant is 
dimensional and consequently it is scale choice dependent. In the standard 
approach, its value is given by $\lambda$ and by the mass standards fixing gauge 
condition (\ref{fix}). Thus we get
\be
\Lambda={\lambda\over 4{\rm !}} ({v\over \chi})^4.
\ee

The very new feature of the Lagrangian (\ref{LT}) is the mixed term (\ref{LSB}) 
that leads to an interaction of Weyl and $U(1)$ Weinberg-Salam vector 
potentials. At quantum level it would result in a mixing of Weyl boson with 
photon and weak bosons - the effect in a sense similar to the known $\gamma - Z$ 
mixing. As the mass of $S_{\mu}$ and the coupling $\alpha_4$ is not predicted by 
the theory, the strength of the mixing effect could be small as well as very 
large. Also the mass $m_S$ cannot be easily estimated from the known data as 
there is no interaction of fermions with the Weyl potential.
Thus definite answers concerning the presence and interactions of Weyl sector 
should be looked in experiments.
\smallskip

{\bf ACKNOWLEDGMENTS}

I am indebted to Professor Jakub Rembieli\'nski, Dr. Kordian Smoli\'nski and 
members of the Organizing Committee for all their efforts in organizing the 
marvelous meeting in \L \'od\'z and for creating the scientific atmosphere. I am 
indebted to Prof. E. Kapu\'scik, Prof. V.N. Pervushin and Dr. A. Horzela for 
valuable discussion. I'm also grateful to Prof. I. Bia\l ynicki-Birula for 
rendering a copy of English translation of \cite{infeld}. The work was supported 
by Polish Committee for Scientific Researches grant no. 2 P03B 183 10.

\end{document}